\documentclass[11pt]{article}
\usepackage{xcolor}
\usepackage{cite}
\usepackage{multirow}
\usepackage{enumerate}
\usepackage[scale=0.8]{geometry}
\usepackage{graphicx}
\usepackage[subrefformat=parens,labelformat=parens]{subfig}
\usepackage{amsfonts,amsmath,amssymb}
\usepackage{multirow,multicol}
\usepackage{bbm}
\usepackage{xspace}
\usepackage{hyperref}
\usepackage{tikz}

\newcommand{\bE}{\ensuremath{\mathbbm{E}}\xspace}
\newcommand{\bR}{\ensuremath{\mathbbm{R}}\xspace}

\begin{document}

~\vspace{.2cm}
\begin{center}
{\LARGE\bfseries Rigor with Machine Learning\\[8pt] from Field Theory to the Poincar\'e Conjecture}\\[1cm]
\textsc{Sergei Gukov}$^{a,b,}$\footnote{gukov@theory.caltech.edu}, \textsc{James Halverson}$^{c,e,}$\footnote{j.halverson@northeastern.edu}, \textsc{Fabian Ruehle}$^{c,d,e,}$\footnote{f.ruehle@northeastern.edu}\\[1cm]
${}^a\;$Richard N. Merkin Center for Pure and Applied Mathematics,\\ California Institute of Technology, Pasadena, CA 91125, USA\\[5pt]
${}^b\;$Dublin Institute for Advanced Studies, 10 Burlington Rd, Dublin, Ireland\\[5pt]
${}^c\;$Department of Physics, Northeastern University, Boston, MA 02115, USA\\[5pt]
${}^d\;$Department of Mathematics, Northeastern University, Boston, MA 02115, USA\\[5pt]
${}^e\;$The NSF AI Institute for Artificial Intelligence and Fundamental Interactions\\[3cm]
\end{center}

\begin{abstract}
Machine learning techniques are increasingly powerful, leading to many breakthroughs in the natural sciences, but they are often stochastic, error-prone, and blackbox. How, then, should they be utilized in fields such as theoretical physics and pure mathematics that place a premium on rigor and understanding? In this Perspective we discuss techniques for obtaining rigor in the natural sciences with machine learning. Non-rigorous methods may lead to rigorous results via  conjecture generation or verification by reinforcement learning. We survey applications of these techniques-for-rigor ranging from string theory to the smooth $4$d Poincar\'e conjecture in low-dimensional topology. One can also imagine building direct bridges between machine learning theory and either mathematics or theoretical physics. As examples, we describe a new approach to field theory motivated by neural network theory, and a theory of Riemannian metric flows induced by neural network gradient descent, which encompasses Perelman's formulation of the Ricci flow that was utilized to resolve the $3$d Poincar\'e conjecture.
\end{abstract}

\thispagestyle{empty}
\clearpage
\tableofcontents

\section{Introduction}

We live in a remarkable time in the history of science: from the perspective of distant descendants, we are near the beginning of the use of computer science for scientific discovery. In the future these discoveries may rely crucially on the promise of quantum computing, but in the present we are faced with an ongoing revolution in artificial intelligence (AI) and machine learning (ML) that is already transforming the natural sciences. See, e.g.\  advances in protein folding \cite{protein_folding} and reviews in the physical sciences \cite{ml_ps} and string theory \cite{Ruehle:2020jrk,He:2023aaa}.

Despite numerous success stories, ML techniques are often stochastic, error-prone, and blackbox. Classification problems always have some degree of error, and this error may be induced by the addition of adversarial noise that, for instance, can cause a neural network to misclassify a turtle as a rifle and a cat as guacamole \cite{athalye2018synthesizing,obfuscated-gradients}. Such ML techniques are stochastic in a number of ways. In the supervised setting that tries to make predictions given model inputs, the predictions of the network, and therefore the error, generally depend on both a random initialization and a random training process. In the deep reinforcement learning setting that tries to learn a strategy for gameplay that maximizes rewards, the training process is stochastic due to randomness in the exploration of the environment. Finally, ML techniques are often blackbox in the sense that the trained ML algorithm has millions or even billions of parameters that are difficult to interpret and understand. These techniques, therefore, may appear ill-suited for application in fields such as theoretical physics and pure mathematics that prioritize rigor and understanding.

\vspace{.2cm}
In this Perspective we survey techniques for obtaining rigor with machine learning, exemplified by recent results. We focus on two central ideas: making applied ML techniques rigorous and ensuring rigor by using theoretical ML from the outset.

We will consider conjecture generation and rigorous solution verification with RL as a means of making applied ML techniques rigorous. In conjecture generation, a human domain expert is brought into the ML loop to understand model predictions with interpretable AI techniques, with the hope of generating a conjecture that can be rigorously proven by a human. We discuss applications of this technique in obtaining new theorems in string theory, algebraic geometry, and knot theory. Rigorous results may also be obtained by reinforcement learning, through the definition of a game whose objective is to find a trajectory through a space of states that establishes a mathematical fact. We will discuss applications of this to string theory and low-dimensional topology. The latter includes a state-of-the-art program that can establish ribbonness of knots (far beyond typical human expert abilities) and which was utilized to rule out hundreds of proposed counterexamples to the smooth Poincar\'e conjecture in four dimensions.

We also review results in an emerging direction that seeks to utilize ML theory in theoretical physics and pure mathematics, including completely new approaches to both field theory and metric flows. These developments utilize recent results on the statistics and dynamics of neural networks. On the statistics side, we present a correspondence between neural networks and quantum field theory (QFT). This correspondence has the potential to provide a non-perturbative definition of QFT in the continuum. We discuss the relation to the Feynman path integral, the neural network origin of interactions and symmetries in a field theory, and the realization of $\phi^4$ theory as a neural network field theory. On the dynamics side, we consider a metric on a Riemannian manifold that is represented by a neural network,  trained with gradient descent. This framework has been used recently in approximations of numerical Calabi-Yau metrics, a crucial recent result in string theory. Neural tangent kernel theory \cite{Jacot2018NeuralTK,widegoogle} is used to develop an associated theory of metric flows that generalizes known flows in mathematics, for instance Perelman's formulation of Ricci flow \cite{Ricci_original} as a gradient flow \cite{perelman2002entropy}, which may be explicitly realized with a neural network metric flow.

There are many research areas and directions not covered in this Perspective, in part due to space limitations. For example, one active area of research is the application of machine learning tools to automated proof assistants. These applications also provide rigorous results, by helping the automated prover to find the right sequence of logical steps to go from statement A to statement B, {\it i.e.} to find a path among logical operations that connects assumptions of a theorem in question to its conclusion \cite{MR3363597}. For use of ML in various systems like Lean, Isabelle, and Mizar see e.g. \cite{MR3153710,MR3543235,HOL,PMA23}.

\section{Rigorous Results from Applied ML}
One way in which ML can lead to rigorous results is by using ML in conjunction with domain experts. An example is conjecture generation, where an ML algorithm assists in formulating a conjecture that can subsequently be proven. Conjectures propose a relation between properties of an object that were  previously not known to be related. Neural networks, being universal function approximators, can be used in a supervised setup to search for such relations: if the algorithm predicts one property from another, this hints at the existence of a relation. If the ML algorithm was a whitebox algorithm like decision trees or symbolic regression, the relation can be inferred directly; in the case of a blackbox algorithm like NNs, attribution techniques can be used. In practice, moving from a trained ML algorithm to a mathematical theorem requires a back-and-forth between the ML and the human to refine the conjecture before it is proven. 

Conjecture generation was introduced in \cite{Carifio:2017bov} in the context of string theory, which used a decision tree and logistic regression, while the use of NNs for this purpose was introduced in~\cite{He:2017aed,Krefl:2017yox,Ruehle:2017mzq}. These ideas were much more recently also applied in knot and graph theory in~\cite{knot_nature,Craven:2020bdz,MR4555584}. Moreover, the utility of performing simple feature reduction techniques like PCA have led to conjectures involving mirror symmetry for Fano varieties~\cite{brown2022computation} or BPS spectra for superconformal field theories (SCFTs) \cite{toappear}. Many more examples can be found in the reviews \cite{Ruehle:2020jrk,He:2023aaa}. See also \cite{mishra2023mathematical} for a recent theoretical approach to conjecture generation.

\medskip

Another way in which ML can lead to rigorous results is to use a trained ML agent to play a game where winning corresponds to solving a scientific problem such that the solution may be rigorously verified. This is the domain of reinforcement learning (RL), which optimizes a policy function that selects actions based on the current state of the system. As the agent explores, it receives rewards that are used to update the policy function, leading to improved behavior over time. RL was famously used in Go and Chess, achieving superhuman performance through self-play, knowing only the rules of the game. These techniques can be interpretable to domain experts via having the trained agent play many games and analyzing the results. For instance in Chess, games played by AlphaZero \cite{Silver2017} demonstrate that it rediscovered the most popular openings used by humans \cite{silver2017mastering}, but also devised useful strategies that surprised grandmasters, such as a proclivity to push the $h$-pawn or favor positional play over pawn-grabbing. We will give examples of both techniques in string theory and knot theory.

\subsection{String Theory and Algebraic Geometry}
Computation of physical quantities in string theory regularly requires algebraic geometry.
The first use of neural networks in string theory targeted questions in computational algebraic geometry~\cite{He:2017aed,Krefl:2017yox,Ruehle:2017mzq}, an area in which  are notoriously computationally complex. They hence provide an interesting target for acceleration via NNs. In all cases, the fact that NNs were able to predict the result with high accuracy suggested the existence of a hitherto unknown relation between the input features and the output. However, extracting an understanding of why the NN manages to perform so well, or whiteboxing the NN, is difficult. Remarkably, the computation of dimensions of line bundle cohomologies, for which~\cite{Ruehle:2017mzq} showed very good performance with a NN, was later discovered~\cite{Klaewer:2018sfl,Brodie:2019ozt,Brodie:2019pnz,Brodie:2019dfx,Brodie:2020wkd} to be related to identifying patterns or regions in the input space, where a simple polynomial related to a topological index describes the individual cohomologies. While this does not prove that the NN in~\cite{Ruehle:2017mzq} made use of this, NNs have been shown to utilize pattern recognition and regression for predictions in other areas of applications. The relation uncovered in~\cite{Krefl:2017yox}, which conjecturally relates the minimum volume of a Sasaki-Einstein cone to simple toric data remains open as of now.

The papers~\cite{Carifio:2017bov, Bies:2020gvf} followed a different route; instead of uncovering an unknown relation by using a NN as a means to check existence of a map from features to labels, the authors use whitebox ML techniques like logistic regression or decision trees from the onset. As the other papers, they target algebraic geometry data. The conjecture uncovered in~\cite{Carifio:2017bov} relates the existence of a ray in a toric variety that defines an elliptically fibered Calabi-Yau fourfold to a particular feature that is interesting for particle physics, $E_6$ gauge symmetry, while~\cite{Bies:2020gvf} is concerned with Brill-Noether theory.

\subsection{Knot Theory and the Poincar\' e Conjecture}
Machine learning is increasingly applied in pure mathematics. We focus on low-dimensional topology, including knots and their relation to the Poincar\'e conjecture. 

\begin{figure}[t]
    \centering
    {
    \fontsize{9pt}{11pt}\selectfont
       \def\svgwidth{5in}
       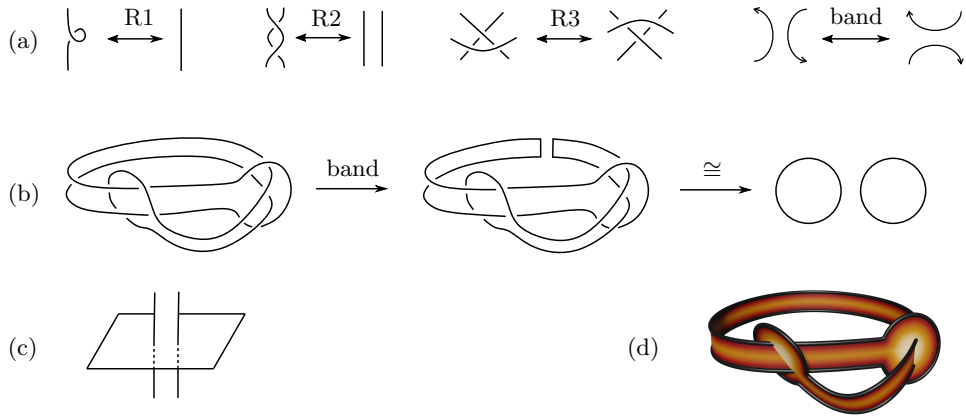
    }
    \caption{(a) The three Reidemeister moves and the band move. The band move needs to preserve orientations on the link. (b) After applying a band move to the square knot, the result can be deformed (via Reidemeister moves) into the unlink with $2$ components. (c) The kind of intersection allowed in a ribbon disk. (d) A ribbon disk for the square knot. Figure taken from~\cite{ribbons}.}
\label{fig:Ribbon}
\end{figure}

A mathematical knot is a circle embedded in $3$-dimensional space; an embedded collection of circles is a generalization known as a link. Its projection to a generic $2$-dimensional plane yields a planar diagram, represented by strands of the knot that cross over or under each other. For a fixed knot or link $K$, the diagram is not unique because $K$ can be continuously deformed and different projections may be chosen. Fortunately, two diagrams represent the same link or knot if and only if they are related by a sequence of simple moves known as Reidemeister moves, shown as the first three moves in Figure \ref{fig:Ribbon} (a). Quantities associated to a knot $K$ that are invariant under Reidemeister moves (or ambient space isotopy) are topological invariants of the knot, and may be integers, vectors, or polynomials, for example.  For works using supervised ML to predict knots invariants, see \cite{MR4101599,Craven:2020bdz,MR4555584}. We will focus on works that obtain rigorous results about knots via conjecture generation or RL.

Conjecture generation was utilized to prove a new theorem about knots \cite{knot_nature}. A fully connected feedforward network was trained to predict an invariant of a knot $K$, known as the signature $\sigma(K)$, from a set $X$ of $12$ knot invariants. The model was trained to high accuracy and gradient saliency via a normalized attribution score  
\begin{equation}
r_i = \frac{1}{|B|} \sum_{x \in X} \left| \frac{\partial L}{\partial x_i}\right|
\end{equation}
was utilized to identify the invariants in $X$ that are most important for predicting $\sigma(K)$. This interpretability analysis identified $3$ invariants (the complex meridional and real longitudinal translations on the boundary torus of the knot complement) as much more significant than the others for determining the knot signature, and they were used by domain experts to generate a conjecture about the signature of a knot. As in the earlier string theory work \cite{Carifio:2017bov}, the conjecture needed refinement by the expert due to counterexamples. A final conjecture was proven, leading to a new theorem that for any hyperbolic knot there exists a constant $c$ such that  
\begin{equation}
|2 \sigma(K) - \text{slope}(K)| \leq \, \text{vol}(K) \text{inj}(K)^{-3},
\end{equation}
in terms of the knot slope, hyperbolic volume, and injectivity radius. Conceptually, the theorem was a surprise to some topologists because it relates geometric topological invariants such as the slope and volume to algebraic topological invariants such as the signature, which a priori seem to be of different character.

\begin{figure}[t]
    \centering
    \includegraphics[width=.99\textwidth]{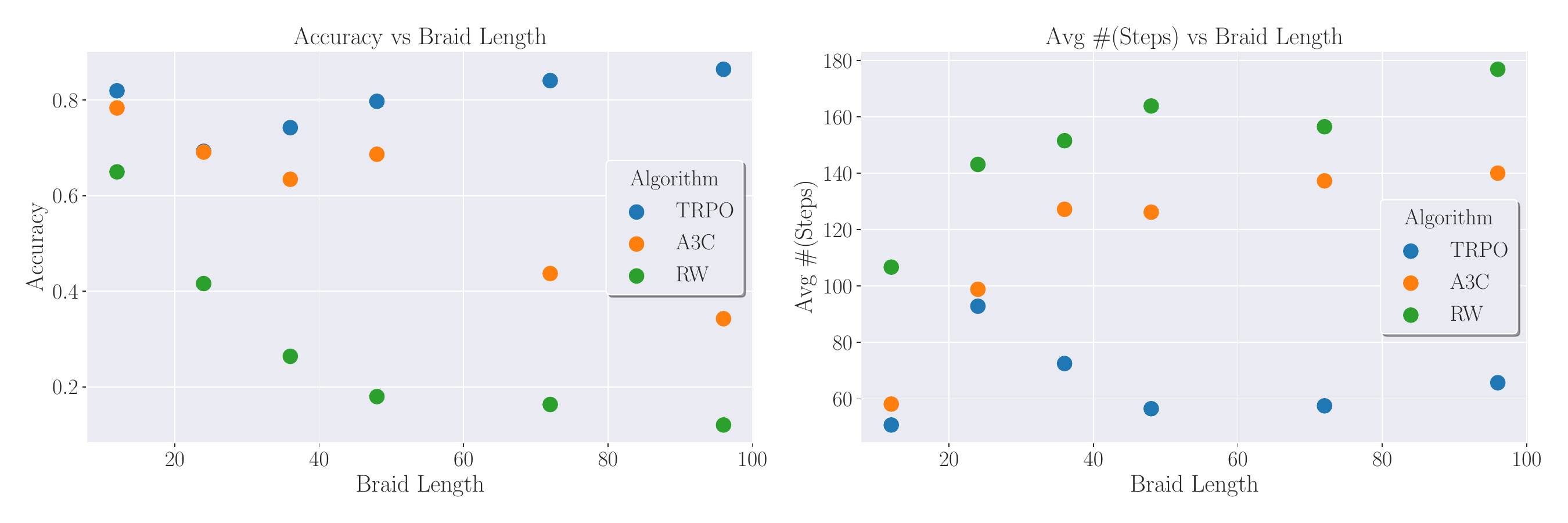}
    \caption{Performance comparison of the TRPO, A3C and RW algorithms on the UNKNOT problem. Left: Fraction of unknots whose braid words could be reduced to the empty braid word as a function of initial braid word length $\ell$. Right: Average number of actions necessary to reduce the input braid word to the empty braid word as a function of $\ell$.}
    \label{fig:RLResults}
\end{figure}

One may also use gameplay to rigorously establish properties of knots. A knot $K$ is said to be the unknot, which is topologically trivial, if there is a sequence of Reidemeister moves that simplifies any planar diagram representing the knot to a standard circle. The UNKNOT problem seeks to determine whether or not $K$ is the unknot, a problem that clearly has complexity growing with the number of crossings $N$ in the planar diagram. The problem is known to be in NP $\cap$ co-NP \cite{MR1693203,MR3177300,lackenby2021efficient}. In \cite{learning}, various RL agents were trained to find sequences of Reidemeister moves that simplify representatives of the unknot to the standard circle. The RL agents significantly outperformed a random walker that selected Reidemeister moves from a uniform distribution. In particular, a trust-region policy optimization (TRPO) agent showed consistent performance for unknots described in terms of braids\footnote{Every knot can be written in terms of an element in a non-Abelian group called the braid group~\cite{Alexander:1923aaa}. The length $\ell$ of corresponding braid word is at least as big as the minimum number $N_\text{min}$ of crossings in the knot projection, $\ell\geq N_\text{min}$.} with increasing number of crossings; see Figure \ref{fig:RLResults}.

A close cousin of the UNKNOT problem is the problem of distinguishing Kirby diagrams that represent 3-manifolds or 4-manifolds. A Kirby diagram basically consists of the data of a link with some integer labels, one for each link component. Much like planar knot projections are related by Reidemeister moves, Kirby diagrams representing the same manifold are related by a small set of moves called Kirby moves. In a simplified setting when all link components are copies of the unknot, this data can be conveniently encoded in a combinatorial structure of a graph, called the plumbing graph. Graph Neural Networks (GNNs) provide a natural choice of architectures to study this problem, on which a trained A3C agent performed extremely well \cite{Ri:2023xcn}, e.g. outperforming the DQN agent.

\bigskip
Another variant of the UNKNOT problem may be utilized to establish that a knot is ribbon, which is related to the smooth $4$-dimensional Poincar\'e conjecture (SPC4). A knot is ribbon if it bounds a ribbon disk, which is one that lives in $3$-dimensional space and has appropriately mild self-intersections. A knot is slice if it bounds a smoothly embedded disk in $4$-dimensional space. Singularities of a ribbon disk are mild enough that there is a canonical procedure to add a dimension and turn it into a smoothly embedded disk in 4d; every ribbon knot is therefore slice. 

The 4D smooth Poincare conjecture (SPC4) states that if a smooth $4$-manifold is homeomorphic to $S^4$, then it is diffeomorphic to $S^4$. SPC4 and sliceness are directly related to one another. Namely, if there is a pair of knots satisfying
\begin{itemize}
\item [a)] $K_1$ and $K_2$ have the same $0$-surgery,
\item [b)] $K_1$ is slice, and
\item [c)] $K_2$ is not slice,
\end{itemize}
then an exotic $4$-sphere may be constructed that is homeomorphic but not diffeomorphic to $S^4$, disproving the long-standing conjecture. Many pairs of knots satisfying a) are known, computation of topological invariants known as slice obstructions may be used to establish c), but there is no known algorithm for establishing sliceness or ribonness. One may address ribbonness by modifying the allowed actions of the UNKNOT problem by adding band addition, see the last move in Figure \ref{fig:Ribbon}. A knot is ribbon if there is a sequence of Reidemeister moves and band additions that simplify a planar diagram representing the knot to a collection of unlinked unknots, see Figure~\ref{fig:Ribbon} (b). 

In \cite{ribbons}, RL and Bayesian optimization of a Markov Decision Process were utilized to establish that a knot is ribbon, via the generalization of the UNKNOT game. Taking speed into account, the strongest overall agent was Bayesian optimized, leading to a state-of-the-art ribbon verifier~\cite{Ruehle:2023Ribbon}, which was able to establish ribbonness of some knots with many crossings (we tested this up to 70 crossings).  The agent was used to demonstrate certain pairs of knots satisfying $a)$ are both ribbon, and therefore both slice. This ruled out over $800$ potential counterexamples to SPC4.

\section{Rigorous Results from ML Theory}

Another option for obtaining rigorous results with machine learning is to use results from ML theory, rather than making numerical results rigorous. This promising approach is still in its infancy, and we will restrict our focus to summarizing work in physics and mathematics that use ML theory results on neural network statistics or learning dynamics.

Central to our discussion will be the use of neural network theory. A neural network is simply a parameterized function
\begin{equation}
\phi_\theta: \mathbb{R}^{n_\text{in}} \rightarrow \mathbb{R}^{n_\text{out}},
\end{equation}
with parameters $\theta$, where one may choose a more general domain and range, if desired. We will often suppress the subscript for notational simplicity. Specifying a neural network requires choosing a concrete functional form, known as the architecture, that is usually a composition of simpler functions. When a neural network is initialized on a computer, however, initial values of the parameters must be chosen by sampling $\theta$ from an initialization parameter distribution $P(\theta)$. The architecture and $P(\theta)$ therefore specify an \emph{ensemble} of functions at initialization, and one question in ML theory regards the statistics of the ensemble.

One aspect of neural network statistics that we will utilize is known as the neural network / Gaussian process (NNGP) correspondence. It was first discovered in the 1990s by Neal~\cite{neal} in the simplest case of a single-layer fully connected width-$N$ neural network, which in the case $n_\text{out}=1$ and no bias in the linear layers has architecture and parameter densities given by
\begin{equation}
\phi(x) = \sum_{i=1}^N \sum_{j=1}^{n_\text{in}} a_i \sigma(w_{ij} x_j)\,, \qquad \qquad P(a)=\mathcal{N}\left(0,\frac{\sigma_a^2}{N}\right)\,, \qquad P(w)=\mathcal{N}\left(0,\frac{\sigma_w^2}{n_\text{in}}\right)\,,
\end{equation}
where $\sigma:\mathbb{R}\to\mathbb{R}$ is an element-wise non-linearity that is part of the architecture choice and the parameter $\theta$ are the matrix components $a_i$ and $w_{ij}$. Neal showed that as $N\to\infty$, $\phi$ is drawn from a Gaussian process. This means that for any set of neural network inputs $\{x_i\}$, the associated vector of outputs $\phi(\{x_i\})$ is drawn from a multivariate Gaussian distribution. Alternatively, in the case of continuous inputs, $\phi(x)$ is drawn from a Gaussian density on functions, with covariance given by the two-point function $G^{(2)}(x,y)= \mathbb{E}[\phi(x)\phi(y)]$ that may be computed \cite{williams} by  integrating over the NN parameters 
\begin{align}
\label{eq:ParameterSpaceDescription}
    G^{(2)}(x,y)= \mathbb{E}[\phi(x)\phi(y)]=\int \text{d}\theta~P(\theta)\, \phi(x)\phi(y)\,.
\end{align}
The corresponding function distribution $P[\phi] := e^{-S[\phi]}$ is specified in terms of the functional
\begin{equation}
S[\phi] = \frac{1}{2} \int \text{d}^{n_\text{in}}x\;\text{d}^{n_\text{in}}y~\phi(x) G^{(2)}(x,y)^{-1} \phi(y),
\label{eqn:NNGP action}
\end{equation}
known as the (Gaussian) action in physics, and $G^{(2)}(x,y)$ is the Feynman propagator. The quantity $G^{(2)}(x,y)^{-1}$ is defined via $\int \text{d}^{n_\text{in}}y\, G^{(2)}(x,y)^{-1} G^{(2)}(y,z) = \delta(x-z)$. 
The NNGP correspondence has been extended to many more architectures \cite{yangTP1}, where the generality of the phenomenon arises from the uniquity of central limit behavior in neural networks, generalizing the notion of ``width.'' The higher moments are correlation functions $G^{(n)}(x_1,\dots,x_n) = \mathbb{E}[\phi(x_1)\dots\phi(x_n)]$ which may also be computed by a parameter space integral akin to~\eqref{eq:ParameterSpaceDescription}, or alternatively via free theory Feynman diagrams with propagator given by $G^{(2)}(x,y)$. Foreshadowing, the NNGP is a generalized free field theory, and we may turn on interactions by including $1/N$-corrections.

Another recent result we will utilize concerns neural network training dynamics. It arises when a neural network is trained by continuous time gradient descent, in which case the dynamics of learning over time $t$ are given by 
\begin{equation}
\frac{\text{d}\phi(x)}{\text{d}t} = \sum_{x'\in D}  \Theta(x,x') \frac{\delta L[\phi(x')]}{\delta \phi(x')}\,, \qquad \qquad 
\Theta(x,x') = \sum_I \frac{\partial \phi(x)}{\partial \theta_I} \frac{\partial \phi(x')}{\partial \theta_I}\,,
\end{equation}
where $\Theta(x,x')$ is called the empirical neural tangent kernel, $D$ is the full training dataset (not a mini-batch), and $L$ is the loss function evaluated on inputs $x'$. This equation arises from a short computation that utilizes only chain rules and the gradient descent update rule $\text{d}\theta_I/\text{d}t$ = $-\sum_{x'\in D}\partial L[\phi(x')]/\partial \theta_I$. The empirical NTK governs the gradient descent dynamics of the finite-width neural network but is difficult to compute since modern neural networks have millions or billions of parameters that evolve in time. However, for many architectures the dynamics simplify in the large-$N$ limit due to a law of large numbers and a natural linearization, yielding 
\begin{equation}
\lim_{N\to\infty} \, \Theta(x,x') =: \bar \Theta(x,x'),
\end{equation}
where $\bar \Theta(x,x')$ is a $t$-independent deterministic kernel that may be computed once-and-for-all at initialization. $\bar \Theta$ may be referred to as the neural tangent kernel (NTK) or the frozen-NTK. If $L$ is mean-squared error loss, the dynamics for all times may be computed exactly analytically \cite{widegoogle} and one may compute the expected prediction of an infinite number of infinite-$N$ neural networks trained to infinite time, allowing a computation that otherwise would not be possible.

\subsection{NN-FT Correspondence}

From the above discussion, we see that neural networks describe ensembles of functions, just like statistical field theories in Physics. In this sense, a neural network architecture and parameter density $P(\theta)$ give a new way to define a field theory.

By the NNGP correspondence, there are many neural network architectures that admit an $N\to \infty$ limit in which the neural network is drawn from a Gaussian process at initialization. Gaussians are determined by the mean and variance, i.e., their one-point correlation function $G^{(1)}(x) = \bE[\phi(x)] = \int \text{d}\theta~P(\theta) \, \phi(x)$ (which is often zero) and two-point correlation function $G^{(2)}(x,y)$ defined in~\eqref{eq:ParameterSpaceDescription}. Any $n$-point correlation function can be computed in parameter space, generalizing two-point calculation~\cite{williams},
\begin{align}
\label{eq:CorrelatorsInParameterSpace}
G^{(n)}(x_1,\dots,x_n)=\bE_\theta[\phi(x_1)\dots\phi(x_n)]=\int \text{d}\theta P(\theta)\phi(x_1)\ldots\phi(x_n)\,,
\end{align}
Crucially, the parameter space integrals may be computable explicitly for some NN architectures, yielding exact correlation functions. In the Gaussian case, $G^{(1)}$ and $G^{(2)}$ determine an associated Gaussian action, which takes the form \eqref{eqn:NNGP action} in the case of vanishing one-point function. Given that action, the correlators may be computed via the Feynman path integral. Hence, the NNGP correspondence is a duality between a Gaussian process and a generalized free field theory. 

This idea can be generalized to a NN-FT correspondence\cite{NNQFT,halverson2021building}, which describes interacting, non-Gaussian theories; see \cite{NNFTs2023} for an introduction and summary of current status. Central to the duality is the idea that field theories may be considered in either the parameter space or function space descriptions. For example, defining a field theory by a NN architecture and parameter density makes the parameter description manifest, but requires some work to determine the associated interacting action $S[\phi]$ that generalizes ~\eqref{eqn:NNGP action} of the NNGP correspondence. Conversely, one could define a field theory in terms of an action $S[\phi]$, as in QFT classes, and then attempt to determine a neural network architecture and parameter density that realizes it. We will see examples of both.

These field theories are usually Euclidean, since most neural networks are defined on $\bR^n$. If the Euclidean theory correlators satisfy the Osterwalder-Schrader axioms \cite{Osterwalder1973}, the theory can be continued to Lorentzian signature, defining a neural network \emph{quantum} field theory \cite{halverson2021building}. For other physics-motivated NN-FT work, see For ML work on finite-width corrections, see e.g. \cite{roberts_yaida_hanin_2022} and references therein, and the introduction of \cite{NNFTs2023} for a discussion of more recent literature.

We emphasize that the duality works in both ways and a theory may be studied in parameter space even if the action is unknown! In other words, the properties of a field theory are determined simply by the architecture and $P(\theta)$. To illustrate this, consider a toy model, where the functions that describe the fields are simply linear functions
\begin{align}
\label{eq:FTLines}
\phi: \bR \to \bR\,,\qquad \phi(x) = \theta x\,,
\end{align}
where the slope $\theta$ is sampled from some distribution $P(\theta)$. The correlators~\eqref{eq:CorrelatorsInParameterSpace} can be computed exactly in this case, $G^{(n)}(x_1,\dots,x_n) = \mu_n \,x_1\dots x_n$, where $\mu_n=\bE_\theta[\theta^n]$ are the moments of $P(\theta)$. The theory has non-Gaussian interactions when $P(\theta)$ is non-Gaussian, meaning its higher cumulants (``connected'' moments) are non-zero for at least some $n>2$. Since Gaussianity of a theory is a consequence of the central limit theorem (CLT), such interactions arise by breaking the assumptions of the CLT: one can either keep $N$ finite or break statistical independence of parameters in $P(\theta)$. Put differently, the change of $P(\theta)$ in parameter space leads to a modification of the function space action that can manifest itself in interaction terms~\cite{halverson2021building,NNFTs2023}. We will see an example for this momentarily.

Before that, let us discuss symmetries~\cite{SymViaDuality} in the NN-FT correspondence, using again the example~\eqref{eq:FTLines}. We see from the exact correlators $G^{(n)}(x_1,\dots,x_n)$ that the theory is scale-invariant, but not translation invariant. If $\mu_{2n+1}=0$, $P(\theta)$ is even and the theory is parity invariant, since the minus sign in $x\to -x$ may be absorbed into a redefinition of $\theta$. Thus, symmetries of the architecture and parameter space density lead to symmetries in function space and vice versa.

To summarize, even without knowing the action, one can
\begin{enumerate}
    \item determine symmetries of the underlying field theory from the architecture and symmetries of $P(\theta)$,
    \item introduce interactions by manipulating $P(\theta)$ or the width $N$ such that it breaks the assumptions of the CLT,
    \item determine interactions by computing parameter space integrals over $P(\theta)$.
\end{enumerate}
In fact, by inverting the correspondence outlined above, one can find the action from the parameter space description of the NN-FT correspondence by computing couplings in terms of Feynman diagrams, whose vertices are the connected correlators~\cite{NNFTs2023}.

To illustrate how a change in $P(\theta)$ turns on interactions, we look at a canonical example in quantum field theory known as $\phi^4$ theory. This theory adds an interaction term $S_\text{int}=\frac{\lambda}{4!}\int\text{d}^{n_\text{in}}x\phi^4(x)$ to the free scalar action $S[\phi] = \int d^{n_\text{in}}x\, \phi(x) \left(\Box + m^2\right) \phi(x)$, a specification of the GP action~\eqref{eqn:NNGP action}. A neural network field theory realizing the free scalar has architecture
\begin{equation}
    \label{eq:phi4}
    \phi_{a,b,c}(x) = \sqrt{\frac{2\, \text{vol}(B^d_\Lambda)}{\sigma_a^2(2\pi)^d}}\,\,\,\sum_{i,j} \frac{a_i \,\cos(b_{ij} x_j + c_i)}{\sqrt{\textbf{b}_i^2 + m^2}}\,.
\end{equation}
with parameters drawn from
\begin{align}
    P_G(a) = \prod_i e^{-\frac{N}{2\sigma_a^2} a_i a_i}\,,\qquad
    P_G(b) = \prod_i P_G(\textbf{b}_i)\,,\qquad
    P_G(c) = \prod_i P_G(c_i)\,,
\end{align}
Here, $P_G(\textbf{b}_i)$ and $P_G(c_i)$ are uniform distributions over an $n_\text{in}$-dimensional ball of radius $\Lambda$ and the interval $[-\pi,\pi]$, respectively. It is perhaps not surprising the the architecture is a normalized plane wave, though other NNGP realizations of the free scalar might exist.
Adding the $\lambda$-dependent term $S_\text{int}$ to the function distribution $P[\phi]$ corresponds, in the parameter space description, to a $\lambda$-dependent deformation of the parameter distribution 
\begin{align}
    \qquad \qquad P(\theta) = P(a,b,c) = P_G(a) P_G(b) P_G(c)\, \,e^{-\frac{\lambda}{4!} \int d^d x\, \phi_{a,b,c}(x)^4}\,,
\end{align}
which is a function purely of parameters via the insertion of the architecture equation \eqref{eq:phi4}.
The existence of interactions in the NN picture arises due to the breaking of statistical independence via the parameter distribution deformation, violating the assumption of independence in the CLT. The result is standard Euclidean $\phi^4$ theory with cutoff $\Lambda$. This NN-FT recovers the NNGP of~\eqref{eqn:NNGP action} in the $\lambda \to 0$ limit~\cite{halverson2021building}, which then has only a single non-vanishing moment,
\begin{equation}
G^{(2)}(p) = \frac{1}{p^2+m^2}\,,
\end{equation}
i.e., $\lambda \to 0$ recovers a free scalar with mass $m$. 
\subsection{Metric Flows with  Neural Networks and the Ricci Flow}

If a neural network is used to represent a metric $g_{ij}$ on a Riemannian manifold $M$, then the training dynamics of the neural network corresponds to a flow in the space of Riemannian metrics. We wish to characterize this flow, following\cite{halversonruehle_flows}.

This idea is motivated in part by recent results on numerical metrics on Calabi-Yau (CY) manifolds. A CY manifold is a complex K\"ahler manifold with vanishing first Chern class $c_1(TM)=0$, where $TM$ is the tangent bundle. By Yau's theorem~\cite{Yau:1978cfy}, any CY manifold with fixed K\" ahler class admits a Ricci-flat K\" ahler metric known as the CY metric, which is unique by a theorem~\cite{Calabi+2015+78+89} of Calabi. 

The proof is non-constructive, however, and \emph{zero} non-trivial explicit CY metrics are known for compact $M$, even though CY manifolds have been studied by mathematicians and string theorists for decades. These facts motivate the use of numerical techniques, e.g.\ via Donaldson's algorithm~\cite{donaldson2005numerical}, or more recently via neural networks that represent the metric \cite{Anderson:2020hux,Douglas:2020hpv,Jejjala:2020wcc,Larfors:2021pbb,Larfors:2022nep,Gerdes:2022nzr}, which give the current state-of-the-art results. For instance, one could train the neural network $g_{ij}$ to minimize a loss given by 
\begin{equation}
L[g] = \sum_{i,j} \left| R_{ij}(g) \right|^2,
\end{equation}
which would drive $g_{ij}$ toward the Ricci-flat metric. This is pedagogically clear, but in practice it is more efficient to use surrogate losses that build in more structure of the problem, for instance rephrasing the problem as a second-order differential equation of Monge-Ampere type~\cite{Yau:1978cfy}.

Returning to flows, if a neural network represents a metric on $M$ and it is trained to approximate the CY metric, then the CY metric is a fixed point of the neural network metric flow. However, another famous metric flow is the Ricci flow \cite{Ricci_original} given by 
\begin{equation}
\frac{\text{d}g_{ij}(x)}{\text{d}t} = -2 \, R_{ij}(x).
\end{equation}
The CY metric is also a fixed point of the Ricci flow, since it is Ricci-flat, $R_{ij}(x)=0$. 
Are these two flows related? A priori, they look very different, since the neural network metric flow is defined by gradient descent on a scalar loss functional (gradient flow) and the Ricci-flow has a tensorial metric update that is not obviously a gradient flow. However, in his work that proved the 3D Poincar\'e conjecture, Perelman showed that a $t$-dependent diffeomorphism of the Ricci flow is a gradient flow defined by 
\begin{equation} 
\frac{\text{d}g_{ij}(x)}{\text{d}t} = \frac{\delta F[\phi,g]}{\delta g_{ij}(x)} = -2 [R_{ij}(x) + \nabla_i \nabla_j \phi(x)] \qquad \qquad F[\phi,g] = \int_M (R+|\nabla \phi|^2)\, e^{-\phi} dV
\end{equation}
where a dilaton field $\phi(x)$ is introduced that has its own dynamics dictated by $\delta F/\delta\phi$. Since this version of the Ricci flow is a gradient flow, it might be possible to realize it as a neural network metric flow.

Motivated by these results, \cite{halversonruehle_flows} developed a theory of neural network metrics flows, assuming that 
a neural network represents a metric $g_{ij}$ on $M$ and the metric is trained by updating the parameters via gradient descent with respect to a scalar loss functional $L[g]$. The dynamics of the metric at $x$ may be computed either from a finite set of points $\{x_i\}$ sampled according to some measure on $M$, or in the continuum (infinite-data) limit. In the continuum case, the metric flow is given by  
\begin{equation}
    \frac{\text{d}g_{ij}(x)}{\text{d}t} = - \int_X \text{d}\mu(x')\,  \Theta_{ijkl}(x,x')  \frac{\delta L[g(x')]}{\delta g_{kl}(x')} \qquad \qquad     \Theta_{ijkl}(x,x') := \sum_I \frac{\partial g_{ij}(x)}{\partial\theta_I} \frac{\partial g_{kl}(x')}{\partial \theta_I}, 
\end{equation}
where $\Theta_{ijkl}(x,x')$ is the empirical metric-NTK that is derived in the same way as the NTK, but keeps track of the tensorial indices associated to the metric. In practice, a natural choice is to take the measure to be the volume measure, $\text{d}\mu = \text{d}V$. This general equation for neural network metric flows is markedly different from Perelman's formulation of Ricci-flow. Unlike the Ricci flow, a general neural network metric flow is governed by the kernel $\Theta_{ijkl}$ that 1) changes the nature of the update equation as it evolves; 2) is non-local, communicating loss fluctuations at $x'$ to metric updates at $x$; and 3) mixes components of the metric in a non-trivial way due to the tensor indices. 

To obtain Perelman's Ricci flow, we must fix the kernel in time, induce locality, and eliminate component mixing. To fix the kernel in time, we take an $N\to \infty$ limit in which the NTK becomes frozen; see e.g. \cite{YangTP2} for a plethora of simple architectures that do this. In such a case the dynamics becomes 
\begin{equation}
    \frac{\text{d}g_{ij}(x)}{\text{d}t} = - \int_X \text{d}\mu(x')\,  \bar \Theta_{ijkl}(x,x')  \frac{\delta L(x')}{\delta g_{kl}(x')} 
\end{equation}
where $\bar \Theta_{ijkl}$ is the metric-NTK in the frozen $N\to \infty$. We call these dynamics an infinite neural network metric flow. Architectures for which the frozen metric-NTK is of the form $\bar \Theta_{ijkl}(x,x') = \overline\Omega(x)\, \delta(x-x') \, \delta_{ik} \delta_{jl}$  for some function $\overline\Omega$ eliminate non-locality and component mixing in the kernel, simplifying the dynamics to 
\begin{equation}
\frac{\text{d}g_{ij}(x)}{\text{d}t} = - \overline\Omega(x) \frac{\delta L[g(x)]}{\delta g_{ij}(x)}.
\end{equation}
We call these dynamics a local neural network metric flow. For any architecture giving such a flow (for which an explicit example is presented in \cite{halversonruehle_flows}), Perelman's formulation of the Ricci flow is obtained simply by choosing the loss to be $L[g(x)] = F[g] / \overline\Omega(x)$. 

We see that NTK theory gives a natural characterization of metric flows induced by neural network gradient descent, demonstrating that they are a significant generalization of the Ricci flow. In particular, for finite neural networks the metric evolves according to a non-local time-dependent kernel that mixes components.

\subsection{Renormalization Group Flows, Optimal Transport, and Bayesian Inference}
ML theory has also begun to interface with physics in connections between renormalization, inference, and diffusion.

Diffusion models have become popular generative models in recent years. For example, to generate one-megapixel images of galaxies, one can draw randomly from a 3M-dimensional pixel space (3M because there are 3 colors per pixel), subject to the constraint that the drawn pixels lead to an image of a galaxy; this means drawing from a complicated distribution in pixel space to achieve that. Diffusion models go the opposite route: starting from images of galaxies, the pixels undergo diffusion and mix until they look like random noise. Inverting this process then leads to a map from noise to images of galaxies.
In terms of probability distributions, one can think of the generation process or the inverse diffusion process as changing the distribution from seemingly random pixels to galaxy picture pixels and the flow in distribution space as an (optimal) transport problem.

From a physics perspective, one can think of diffusion as ``destroying'' information in the image; removing data points from an inference problem has a similar effect. As a process, this resembles renormalization group flow in quantum field theory, where information is lost through coarse-graining which forgets about irrelevant operators. Keeping track of these operators leads to the notion of an exact renormalization group flow. The authors of ~\cite{Cotler:2023,Berman:2022uov} used this to connect the exact renormalization group flow to a optimal transport and Bayesian inference, respectively. Phrased in terms of Bayesian inference, where the information is added through updating the prior as new observations come in, this allows to give an information-theoretic meaning to exact renormalization.

\section{Outlook}

In this Perspective, we have reviewed applications of ML to theoretical physics and pure mathematics. These fields are late adopters of ML, since they require rigor and interpretability, while traditional ML techniques are often stochastic, error-prone, and blackbox. Application of ML techniques to these fields thus require rethinking and modifying techniques that are readily applied in other natural sciences. We focused on two main avenues: making applied ML rigorous, and applying theoretical ML. We exemplified the former with conjecture generation and rigorous verification by reinforcement learning, and the latter using neural network theory.

In conjecture generation, a human is brought into the loop to interpret what the ML algorithm has learned and turn it into a conjecture which can then be refined and proven by domain experts. This has been done in supervised ML: if a NN or other algorithm (e.g., symbolic regression or decision trees) can learn a high-accuracy map from inputs to the labels, for which no known relation exists, this hints at a new connection. Given that setting up such supervised learning problems can be done very quickly, this allows for scanning theoretical data for new relations.  We gave examples of this idea in string theory, algebraic geometry, and knot theory. 

A second avenue is to use reinforcement learning for problems in mathematics. The idea is not to attempt to whitebox the ML algorithm, but instead to look at episode rollouts of the RL agent to infer the solution strategy learned by the agent. In particular, RL rollouts can be used to obtain (provably correct) truth certificates for decision problems of the type ``Does object O have property X?'' RL is useful in such cases since we can set up an RL algorithm that manipulates object O until property X is manifest and the algorithm reaches a terminal state. The rollouts of the episode that lead to the terminal state are then the truth certificate for the decision problem. Rollouts were used to rule out hundreds of proposed counterexamples to the smooth Poincare conjecture in 4 dimensions, and to establish sliceness for new knots. 

A different approach is to use ML theory to obtain rigorous results, entirely avoiding error introduced by applied ML techniques. For instance, due to the central limit theorem, the statistics of the functions expressed by certain NNs become tractable in the infinite parameter regime; they are draws from Gaussian processes. In the context of Physics, this leads to a correspondence between NN and statistical field theories, where the infinite parameter regime defines generalized free field theories and leaving this regime corresponds to turning on interactions. This provides a new \emph{definition} of a field theory, motivated by ML theory, and therefore opens a new approach for the study of new and existing field theories. We exemplified a few aspects of the correspondence, including the role of the central limit theorem, the origin of symmetries, and the realization of $\phi^4$ theory as a neural network field theory. Hopefully, in the future it can lead to better understanding of non-perturbative QFTs, which at present is one of the major open problems in field theory.

We also reviewed the theory of flows in the space of Riemannian metrics induced by gradient descent when the metric is modeled as a neural network. This theoretical framework encompasses, for instance, recent empirical results that use neural networks as state-of-the-art approximations to Calabi-Yau metrics.  In the infinite parameter limit the metric flow simplifies and neural tangent kernel theory may be utilized. Under some additional architecture assumptions, the flow may be made local, and Perelman's famous formulation of Ricci flow as a gradient flow is realized as a neural network metric flow.

\subsection*{Acknowledgements}
S.G.\ is supported in part by a Simons Collaboration Grant on New Structures in Low-Dimensional Topology and by the DOE grant DE-SC0011632. J.H.\ is supported by NSF CAREER grant PHY-1848089. F.R.\ is supported by the NSF grants PHY-2210333 and startup funding from Northeastern University. The work of J.H.\ and F.R.\ is also supported by the National Science Foundation under Cooperative Agreement PHY-2019786 (The NSF AI Institute for Artificial Intelligence and Fundamental Interactions). 

\bibliographystyle{bibstyle}
\bibliography{refs}

\end{document}